\documentclass[a4paper,11pt]{article}

\usepackage{contribution}

\newcommand{\weblink}[2][]{%
    \ifthenelse{\equal{#1}{}}%
    {\textnormal{\url{#2}}}%
    {\textnormal{\href{#2}{#1}}}%
}

\newcommand{\acknowledgements}[1]{%
  \bigskip\bigskip
  \textsf{\textbf{\Large Acknowledgements}} \\[2ex]
  {#1}
  \bigskip
}

\def\beq{\begin{equation}}
\def\eeq#1{\label{#1}\end{equation}}
\def\eeqn{\end{equation}}

\def\beqa{\begin{eqnarray}}
\def\eeqa#1{\label{#1}\end{eqnarray}}
\def\eeqan{\end{eqnarray}}

\let\bar=\overbar

\def\Dslash{\not{\hbox{\kern-4pt $D$}}}
\def\dslash{\not{\hbox{\kern-2pt $\del$}}}

\def\msb{{\bar{\ssstyle M \kern -1pt S}}}

\newcommand{\contribution}[7][]{%
  \clearpage
  \thispagestyle{plain}
  \ifthenelse{\equal{#1}{}}
  {\hypersetup{pdftitle={#2}}}
  {\hypersetup{pdftitle={#1}}}
  \hypersetup{pdfauthor={{#3} {#4}}}
  {\centering\normalfont\LARGE\bfseries\sffamily #2 \par\nobreak}
  \lhead{}
  \chead{%
    \textit{\footnotesize XIV International Conference on Hadron Spectroscopy
      (\weblink[\textit{hadron2011}]{http://www.hadron2011.de}), 13-17 June 2011, Munich, Germany}%
  }
  \rhead{}
  \bigskip
  \begin{center}
    {#3} {#4}\ifthenelse{\equal{#6}{}}{}{\footnote{\weblink[#6]{mailto:#6}}}
    \ifthenelse{\equal{#7}{}}{}{#7} \\
    \textit{#5}
  \end{center}
  \bigskip
}

\renewcommand{\abstract}[1]{%
  \begin{center}
    \begin{minipage}{0.85\textwidth}
      \begin{footnotesize}
        #1
      \end{footnotesize}
    \end{minipage}
  \end{center}
  \bigskip
}

\begin{document}

%
%
%
%
%
{  

\makeatletter
\@ifundefined{c@affiliation}%
{\newcounter{affiliation}}%
{}%
\makeatother
\newcommand{\affiliation}[2][]{\setcounter{affiliation}{#2}%
  \ensuremath{{^{\alph{affiliation}}}\text{#1}}}
%
%

\contribution[Kaonic $^3$He and $^4$He X-ray measurements in SIDDHARTA] 
{Kaonic $^3$He and $^4$He X-ray measurements in SIDDHARTA}  
{T.}{Ishiwatari}  
{\affiliation[Stefan-Meyer-Institut f\"{u}r subatomare Physik, Vienna, AUSTRIA]{1}\\
 \affiliation[INFN, Laboratori di Frascati, Frascati (Roma), ITALY]{2}\\
 \affiliation[Dep. of Phys. and Astro., Univ. of Victoria, Victoria B.C., CANADA]{3}\\
 \affiliation[Politecnico di Milano, Sez. di Elettronica, Milano, ITALY]{4}\\
 \affiliation[IFIN-HH, Magurele, Bucharest, ROMANIA]{5}\\
 \affiliation[INFN, Sez. di Roma I and Inst. Superiore di Sanita, Roma, ITALY]{6}\\
 \affiliation[Univ. of Tokyo, Tokyo, JAPAN]{7}\\
 \affiliation[RIKEN, The Inst. of Phys. and Chem. Research, Saitama, JAPAN]{8}\\
 \affiliation[Tech. Univ. M\"{u}nchen, Physik Dep., Garching, GERMANY]{9}
}
{tomoichi.ishiwatari@assoc.oeaw.ac.at}  
{\!\!$^,\affiliation{1}$, 
M.~Bazzi\affiliation{2}, 
G.~Beer\affiliation{3}, 
C.~Berucci\affiliation{2}, 
L.~Bombelli\affiliation{4}, 
A.M.~Bragadireanu\affiliation{2}\ensuremath{^,}\affiliation{5},
M.~Cargnelli\affiliation{1}, 
A.~Clozza\affiliation{2}, 
G.~Corradi\affiliation{2}, 
C.~Curceanu (Petrascu)\affiliation{2}, 
A.~d'Uffizi\affiliation{2}, 
C.~Fiorini\affiliation{4}, 
F.~Ghio\affiliation{6}, 
B.~Girolami\affiliation{6}, 
C.~Guaraldo\affiliation{2}, 
R.S.~Hayano\affiliation{7}, 
M.~Iliescu\affiliation{2}\ensuremath{^,}\affiliation{5},
M.~Iwasaki\affiliation{8}, 
P.~Kienle\affiliation{9},
P.~Levi Sandri\affiliation{2}, 
V.~Lucherini\affiliation{2}, 
J.~Marton\affiliation{1}, 
S.~Okada\affiliation{2}, 
D.~Pietreanu\affiliation{2}, 
K.~Piscicchia\affiliation{2}, 
M.~Poli~Lener\affiliation{2}, 
T.~Ponta\affiliation{5}, 
R.~Quaglia\affiliation{4}, 
A.~Rizzo\affiliation{2}, 
A.~Romero Vidal\affiliation{2}, 
A.~Scordo\affiliation{2}, 
H.~Shi\affiliation{7}, 
D.L.~Sirghi\affiliation{2}\ensuremath{^,}\affiliation{5},
F.~Sirghi\affiliation{2}\ensuremath{^,}\affiliation{5},
H.~Tatsuno\affiliation{2},
A.~Tudorache\affiliation{5}, 
V.~Tudorache\affiliation{5}, 
O.~Vazquez Doce\affiliation{2},
E.~Widmann\affiliation{1},
B.~W\"{u}nschek\affiliation{1},
J.~Zmeskal\affiliation{1},
}  
%

\abstract{%
The strong-interaction shift of kaonic $^3$He and $^4$He $2p$ states
was measured using gaseous targets for the first time in 
the SIDDHARTA experiment. The determined shift of kaonic $^4$He is much smaller than
the values obtained in the experiments performed in 70's and 80's. Thus, the problems
in kaonic helium (the ``kaonic helium puzzle'') was definitely solved by our measurements.
The first observation of the kaonic $^3$He X-rays was also achieved. 
The shift both of kaonic $^3$He and $^4$He was found to be as small as a few eV.
}
%

\section{Introduction}
There was a serious problem in the strong-interaction shift of the kaonic $^4$He $2p$ level.
The experimentally determined shift of the kaonic $^4$He $3d \to 2p$ X-ray transition
gave a large shift (about $-40$ eV) \cite{Baird}, while a predicted value deduced from 
optical models using the kaonic atom data of $Z \ge 3$ was about $0$ eV \cite{Batty}.
Recently this abnormal $2p$ level shift was focused on  theoretical studies related to 
kaonic nuclear states, where a shift as large as 10 eV was estimated \cite{Akaishi}. 
However, no theory could obtain such a large measured shift of -40 eV. Therefore, 
confirmation measurements were awaited for a long time.

The experiment by the KEK E570 collaboration determined the shift of 
$+2 \pm 2 \mbox{ (stat)} \pm 2 \mbox{ (syst)}$ eV \cite{E570}, which disagreed with the 
previously measured value of $-43 \pm 8$ eV.

Therefore, the SIDDHARTA experiment measured the kaonic $^4$He X-rays to confirm the new result
obtained by E570 \cite{He4}. In addition, the kaonic $^3$He X-rays were measured using the 
same experimental apparatus, by replacing the target gas \cite{He3}.  The latter was the 
world first's observation. This experiment provided new results on the kaon-helium interaction.

\section{Experiment}
The kaonic helium X-rays were measured in the SIDDHARTA experiment, which was 
performed in the DA$\Phi$NE $e^+e^-$ collider at LNF (Italy). Charged kaon pairs 
produced by the annihilation of $e^+ e^-$ beams were detected by two scintillators installed in the
interaction point. X-rays were detected using large area silicon drift detectors (SDDs)
with a total area of 144 cm$^2$ \cite{NIM}. Background events associated with the accelerator
were rejected using the timing between the X-ray hits on the SDDs and the coincidence of $K^+K^-$ pairs.
A gaseous target was used in the measurement. The advantage was the negligible effect of 
Compton scattering in helium, where  the contribution of the Compton tail was one 
of the difficulties in the analysis in the previous experiments \cite{Baird, E570}.
\begin{figure}[htb]
  \begin{center}
    \includegraphics[width=0.45\textwidth]{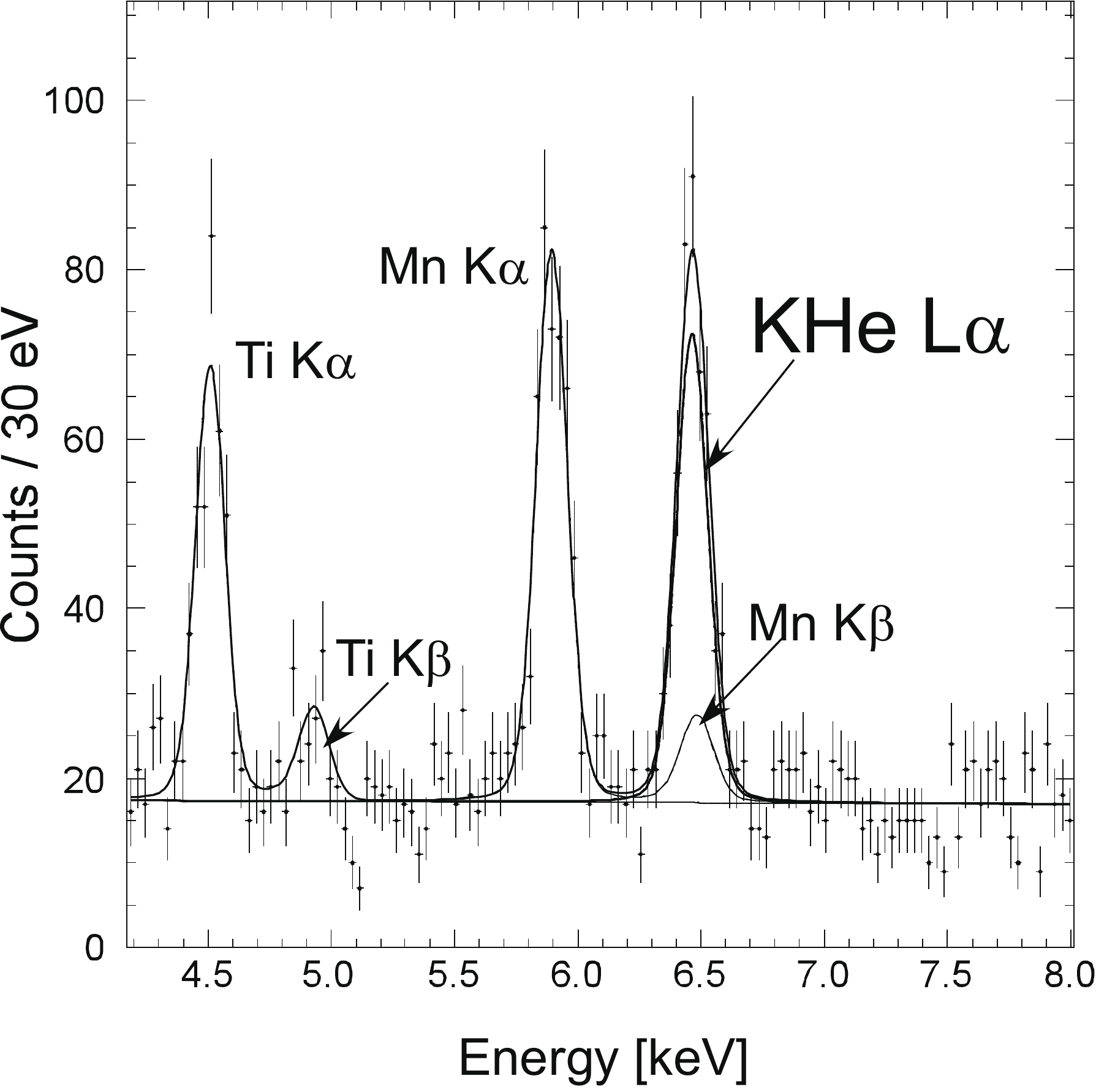}
    \includegraphics[width=0.45\textwidth]{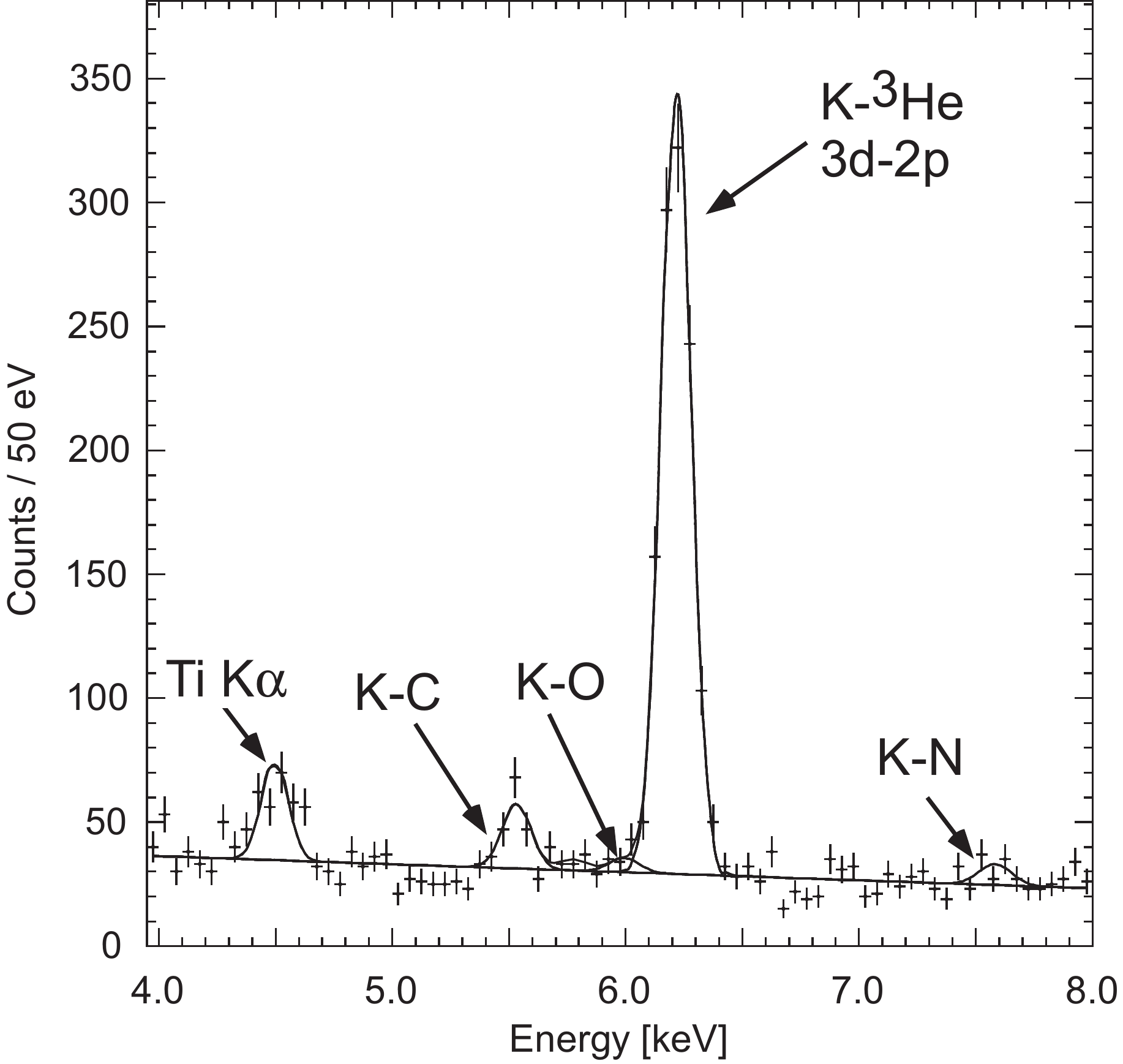}
    \caption{X-ray energy spectra of the $3d \to 2p $ transition of 
kaonic $^4$He (left) and $^3$He (right). The energy of these transitions were determined
within an accuracy of a few eV.}
    \label{fig:spec}
  \end{center}
\end{figure}

The data of kaonic helium were taken during the beam time in 2009.
The X-ray energy data of each SDD were calibrated using the X-ray tube
installed in the setup. The background events uncorrelated to the timing of 
production of kaonic atoms were rejected by the analysis. A detailed description of the
analysis can be found in \cite{He3,He4,NIM}.

Figure ~\ref{fig:spec}(left) shows the X-ray energy spectrum of kaonic $^4$He \cite{He4}. 
The position of the kaonic helium X-rays are indicated in the figure. The Mn K$\alpha$ 
and K$\beta$ lines were originated from an $^{55}$Fe source installed in the
setup. These lines were used for calibration purpose and stability check of the SDDs \cite{NIM}.
The right figure shows the X-ray energy spectrum of kaonic $^3$He. In addition to the kaonic
$^3$He X-ray line, some kaonic atom X-ray lines were seen, which were originated from the material
(Kapton Polyimide) of the target cell window \cite{He3}.

The strong interaction shift of the kaonic He $2p$ states was obtained
from the fit of the kaonic He X-ray peaks. The determined shifts are
shown in Fig.~\ref{fig:results}.


\section{Conclusions}

The SIDDHARTA experiment measured the strong-interaction shift
both of the kaonic $^3$He and $^4$He $2p$ levels with an accuracy of several eV.
They were measured using gaseous targets for the first time, and the world first's observation
of kaonic $^3$He X-rays was achieved.

A large shift of the order of $-40$ eV determined by the experiments
performed in 70's and 80's was not obtained neither in kaonic $^3$He nor kaonic $^4$He. 
Both are consistent with 0 eV within the errors.
The results agree with theoretical predictions determined from  
other kaonic atoms with $Z \ge 3$ using optical model approaches \cite{Batty}. However,
the SIDDHARTA results may indicate a possible isotope shift between kaonic $^3$He and kaonic $^4$He.
Also, the theoretical prediction by \cite{Akaishi} cannot be excluded within our accuracy.

More precise measurements are very important to understand the
kaon-helium interaction. Such precision measurements can be performed 
in the J-PARC E17 experiment \cite{E17}.

\begin{figure}[htb]
  \begin{center}
    \includegraphics[width=0.4\textwidth]{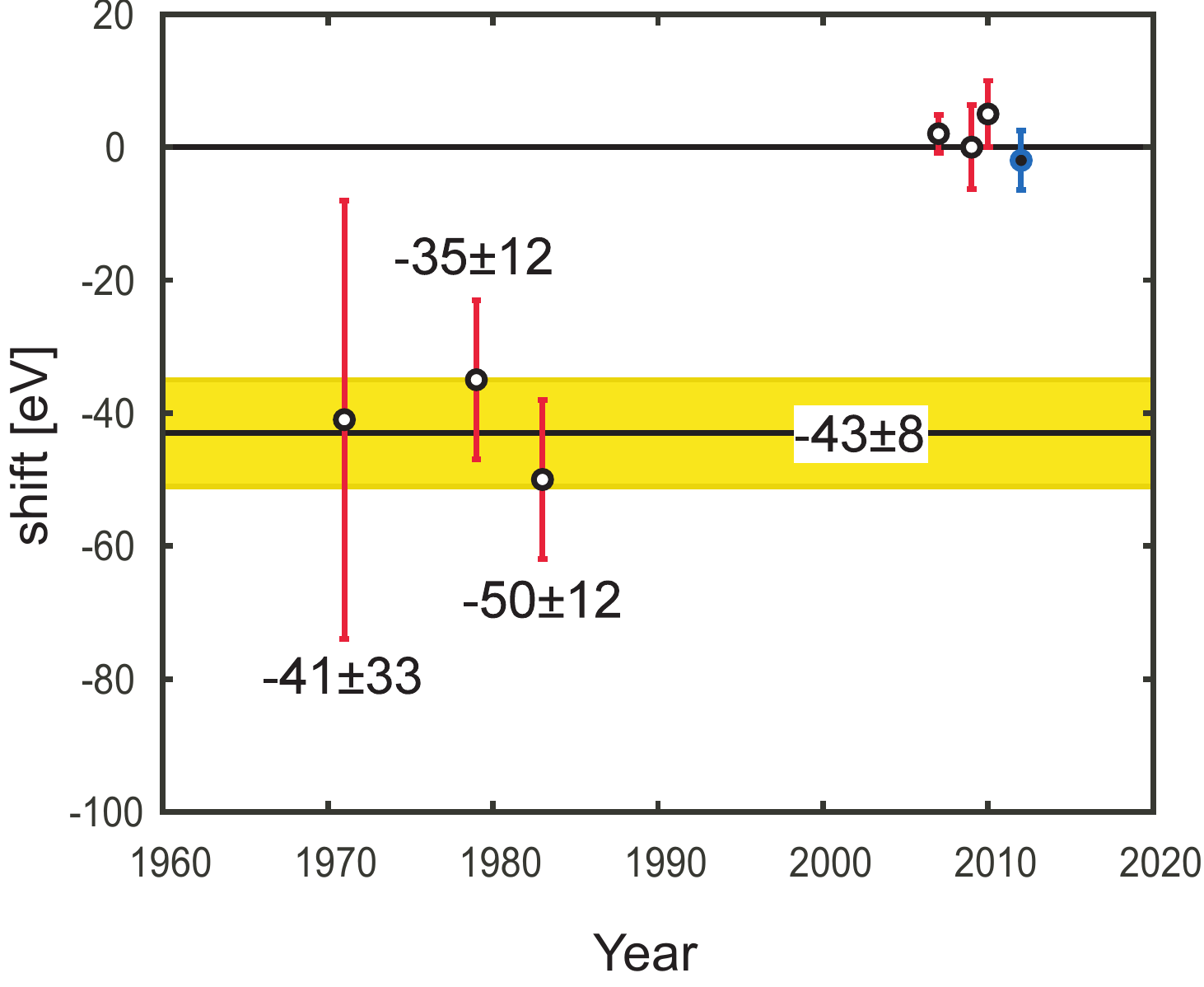}
    \includegraphics[width=0.4\textwidth]{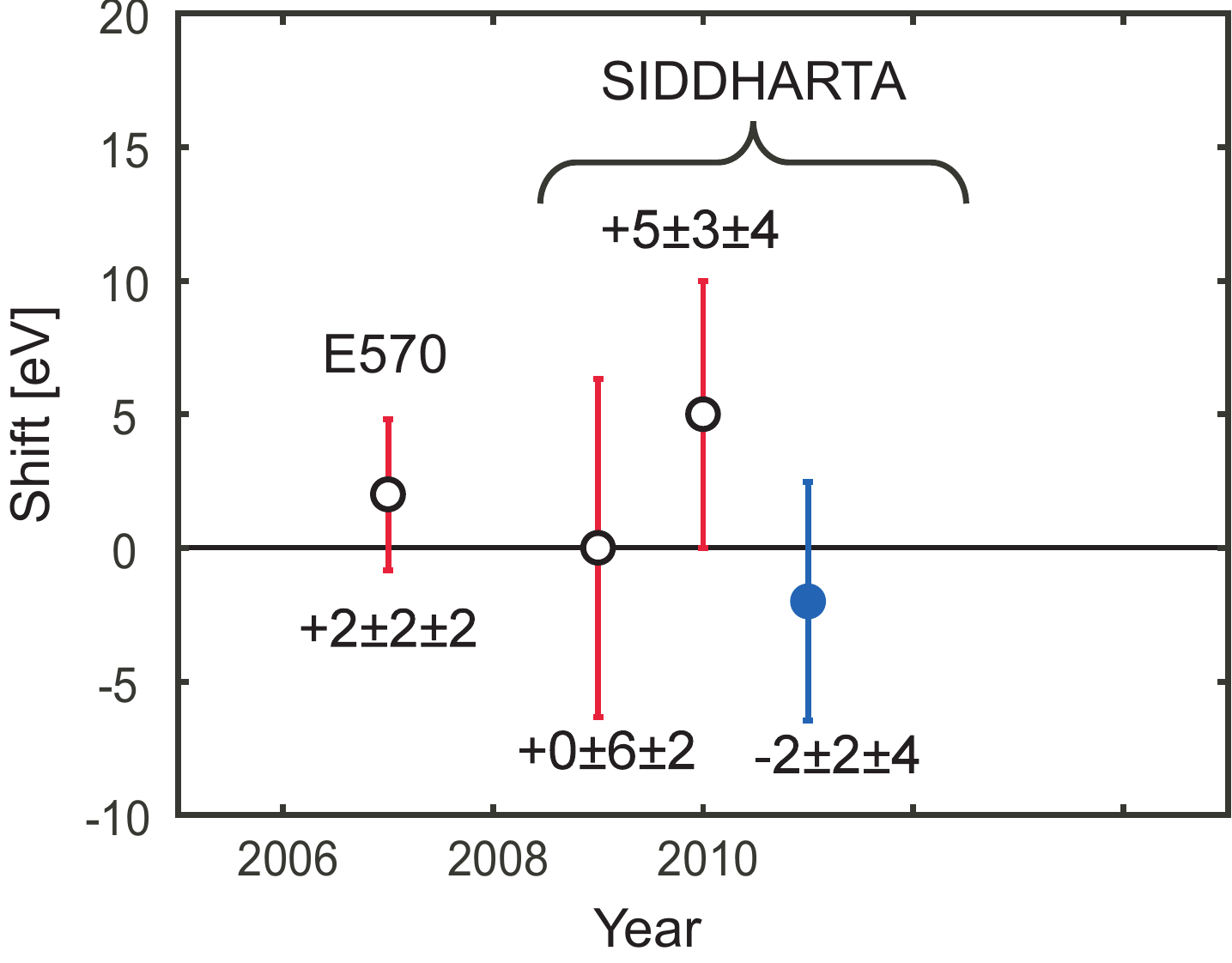}
    \caption{Comparison of experimental results. 
Open circle: K-$^4$He $2p$ state, Closed circle: K-$^3$He $2p$ state. A negative (positive)
value of the shift shows a repulsive (attractive) shift.}
    \label{fig:results}
  \end{center}
\end{figure}

\acknowledgements{%
We thank C. Capoccia, B. Dulach, and D. Tagnani from
LNF-INFN; and H. Schneider, L. Stohwasser, and  D. St\"{u}ckler
from Stefan-Meyer-Institut,
 for their fundamental contribution in designing and building
the SIDDHARTA setup.
We thank as well the DA$\Phi$NE staff for the excellent working
conditions and permanent support.
Part of this work was supported by HadronPhysics I3 FP6 European
Community program, Contract No. RII3-CT-2004-506078; 
the European Community Research Infrastructure Integrating 
Activity ``Study of Strongly Interacting Matter'' 
(HadronPhysics2, Grant Agreement No. 227431) under
the Seventh Framework Programme of EU;
Austrian Federal Ministry of Science and Research BMBWK  
650962/0001 VI/2/2009; Romanian National Authority for Scientific Research, 
Contract No. 2-CeX 06-11-11/2006; Grant-in-Aid for Specially 
Promoted Research (20002003), MEXT, Japan; and the Austrian Science
Fund (FWF): [P20651-N20].
}


%

}  



\begin{thebibliography}{99}
  
\bibitem{Baird} 
S. Baird {\it et al.}, Nucl. Phys. A {\bf 392}, 297 (1983).

\bibitem{Batty}
C.J. Batty, Nucl. Phys. A {\bf 508}, 89c (1990).

\bibitem{Akaishi}
Y. Akaishi, {\it Proc. Inter. Conf. on Exotic Atoms (EXA05)}, 
Austrian Academy of Sciences Press, Vienna (2005), p. 45;
\url{http://dx.doi.org/10.1553/exa05s45}. 

\bibitem{E570}
S. Okada {\it et al.}, Phys. Lett. B {\bf 653}, 387 (2007).

\bibitem{He4} 
SIDDHARTA Collaboration, Phys. Lett. B {\bf 681}, 310 (2009).

\bibitem{He3}
SIDDHARTA Collaboration, Phys. Lett. B {\bf 697}, 199 (2011).

\bibitem{NIM}
M. Bazzi {\it et al.}, Nucl. Instr. and Meth. A {\bf 628}, 264 (2011).

\bibitem{E17}
R. S. Hayano {\it et al.}, Proposal of J-PARC 50-GeV PS,
{\it ``Precision spectroscopy of Kaonic Helium 3 $3d \to 2p$ X-rays''},(2006).

\end{thebibliography}
\end{document}